%% file: Arxiv.tex
\documentclass{article}

\input{commands}

\usepackage{tikz,calc}
\usetikzlibrary{fit}
\tikzstyle{point}=[circle,fill,inner sep=1.2pt]
\usepackage[numbers,sort&compress]{natbib}

\title{Flip Distance to a Non-crossing Perfect Matching}

\author{
\'{E}douard Bonnet \thanks{Institute for Computer Science and Control,
Hungarian Academy of Sciences (MTA SZTAKI), 
{\tt edouard.bonnet@lamsade.dauphine.fr}
}
\and
Tillmann Miltzow
\thanks{Institute for Computer Science and Control,
Hungarian Academy of Sciences (MTA SZTAKI), {\tt t.miltzow@gmail.com}
}
}

\begin{document}

\maketitle

\begin{abstract}
 A  \emph{perfect straight-line matching} $M$ on a finite set $P$ of points
in the plane is a set of  segments such that each point in $P$ is an endpoint
of exactly one segment. $M$ is \emph{non-crossing} if no two segments in $M$ cross each other. 
Given a perfect straight-line matching $M$ with at least one crossing, we can remove this crossing by a flip operation. 
The flip operation removes two crossing segments on a point set $Q$ and adds two non-crossing segments to attain a new perfect matching $M'$.
It is well known that after a finite number of flips, a non-crossing matching is attained and no further flip is possible. 
However, prior to this work, no non-trivial upper bound on the number of flips was known.
If $g(n)$ (resp.~$k(n)$) is the maximum length of the longest (resp.~shortest) sequence of flips starting from any matching of size $n$, we show that $g(n) = O(n^3)$ and $g(n) = \Omega(n^2)$ (resp.~$k(n) = O(n^2)$ and $k(n) = \Omega (n)$).
\end{abstract}

\section{Introduction}
\begin{figure}[htbp]
 \centering\includegraphics{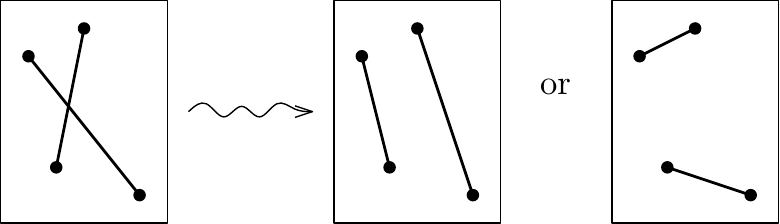}
 \caption{Two crossing segments are replaced by two non-crossing segments. There are two ways to flip.}
 \label{fig:FlipDef}
\end{figure}
\begin{figure}[htbp]
 \centering\includegraphics{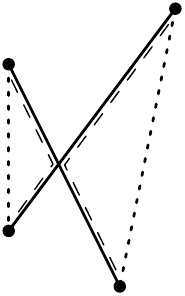}
 \caption{The two new edges (dotted) are shorter than the old edges (solid) since the dashed part to the left (resp.~to the right) of the crossing is longer than the dotted segment on the left (resp.~on the right).}
 \label{fig:Length}
\end{figure}

Given $2n$ points in the plane in general position (no three points on a line), we define a \emph{perfect straight-line non-crossing matching} as a set of $n$ segments such that each point is incident to exactly one segment and no two segments intersect. Given $2n$ points in the plane, it is well-known that a perfect straight-line non-crossing matching always exists. One elegant argument to see this is to start with any perfect straight-line matching, potentially self-intersecting, and remove any crossing by a flip (see Figure~\ref{fig:FlipDef}).
Although the total number of crossings might increase (see Figure~\ref{fig:MoreCrossings}), the sum of the length of all the segments decreases (see Figure~\ref{fig:Length}). 
Thus, the process will eventually end with a perfect non-crossing matching.

A simpler argument is to take the first two points with lowest $x$-coordinate and connect them with a  segment and continue with the remaining points by induction. 
Contrary to the first argument, this does not carry over to the bichromatic setting (where points are partitioned into two color classes and only segment linking points of different colors are allowed).

Motivated by this old folklore result, we investigate the question on the maximum and minimum number of flips that are necessary and sufficient to reach a straight-line non-crossing matching.

\subsection{Preliminaries}
From here on, $P$ always denotes a set of $2n$ points in the plane and $M$ a perfect straight-line matching on $P$.
Given two points $a$ and $b$ we denote by $\mbox{seg}(a,b)$ the segment with endpoints $a$ and $b$.
%
Matching $M$ is a \emph{successor} of matching $M'$ if we can construct $M$ from $M'$ by a single flip. 
We say that $\mathcal{M} = (M_0,\ldots,M_k)$ is a valid sequence of matchings, if each matching $M_{i+1}$  is a successor of $M_i$ and $M_k$ is non-crossing. 
The number $k$ denotes the length of $\mathcal{M}$. 
Given a set $P$ of $2n$ points in the plane, we define: 
\[f(M) = \max \{ \, k : \exists \mathcal{M} \mbox{ of length } k  \mbox{ with } M = M_0\, \} ;\]
\[h(M) = \min \{ \, k : \exists \mathcal{M} \mbox{ of length } k  \mbox{ with } M = M_0\, \}.\]
Consequently functions $g(n)$ and $k(n)$ are defined as:
\[g(n) = \max \{\,  f(M) : M \mbox{ is a matching on } 2n \mbox{ points} \, \};\]
\[k(n)=\max \{ \, h(M) : M \mbox{ is a matching on } 2n \mbox{ points} \, \}.\]


\subsection{Results}
We establish the following result:
\begin{theorem}\label{thm:max} Let $n$ be a large enough natural number then holds:
  \[ \frac{n^2 - n}{2} = \binom{n}{2} \leq g(n) \leq   n^3 .\] 
\end{theorem}

This result immediately carries over to bichromatic matchings.
We conjecture that $g(n) = \Theta(n^2)$.

\begin{theorem}\label{thm:min} Let $n$ be a large enough natural number then holds:
  \[  n-1 \leq k(n) \leq   \frac{n^2}{2} ,\] for some constants $C$. 
\end{theorem}

Our proof of Theorem~\ref{thm:min} does \emph{not} carry over to the bichromatic case. 
However, we will see that the upper bound further holds if the crossing to flip is imposed at each step by an adversary and we may only choose which of the two flips (see Figure~\ref{fig:FlipDef}) we perform.

\subsection{Related Work}

The combinatorial work on flip graphs of geometric structures is fairly large. 
See the survey by Bose and Hurtado~\cite{DBLP:journals/comgeo/BoseH09} for an overview and some motivations.

Matchings, triangulations and spanning trees are commonly studied in recent work~\cite{DBLP:journals/corr/AsinowskiR15,
DBLP:journals/comgeo/AloupisBLS15, asinowski2013quasi,
DBLP:journals/combinatorics/AichholzerAM15, ishaque2013disjoint,
DBLP:journals/dcg/AichholzerMP15, 
DBLP:journals/ijcga/AichholzerHOPSV14, houle2005graphs,
DBLP:journals/gc/AichholzerHORRSS13, 
DBLP:journals/comgeo/AichholzerBDGHHKMRSSUW09, 
DBLP:journals/comgeo/OlaverriHHT14, 
DBLP:journals/ipl/AichholzerAHK06}.
Particularly interesting are triangulations of points that are in convex positions as they correspond to Catalan structures.
Another interesting application comes from Lawson flips, which can be used to reach the Delaunay triangulation in $O(n^2)$ flips~\cite{lawson1986properties}. 
This can be used for the reverse search technique to enumerate triangulations~\cite{avis1996reverse}.

%
%
%
%
%
%
%
%
%
%
%
%
%

\section{Lower Bounds}
\begin{figure}[htbp]
 \centering\includegraphics{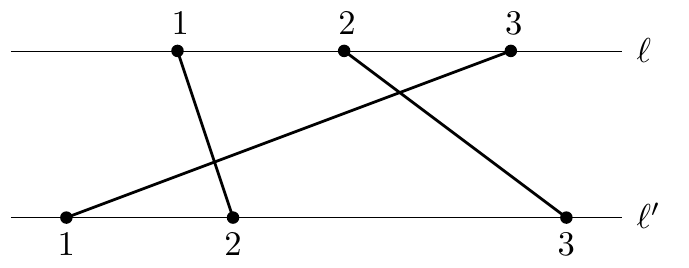}
 \caption{Matching corresponding to cycle $(123)$.}
 \label{fig:LowerBound}
\end{figure}

%

We start with the lower bound for Theorem~\ref{thm:max} and~\ref{thm:min}. Let $\ell$ and $\ell'$ be two parallel horizontal lines and let $P$ be a set of $2n$ points, $n$ of which are on $\ell$ and $\ell'$ respectively.
In the following, we consider only matchings that connect points from $\ell$ to $\ell'$ (see Figure~\ref{fig:LowerBound}).

Every such matching $M$ can be interpreted as a permutation $\pi_M$ and $M$ is crossing free if and only if $\pi_M$ is the identity. 
We can always do flips that correspond to an elementary step in bubble sort. 
Bubble sort on permutation $\pi$ needs as many steps as the number of inversions of $\pi$. 
And, the number of inversions is at most $\binom{n}{2}$. 
A small perturbation of the point set ensures general position.

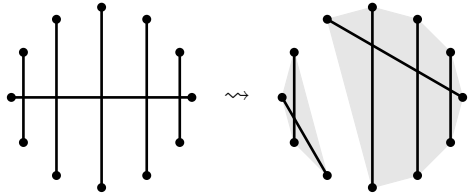
\begin{figure}[h]
\centering
\begin{tikzpicture}
\def\n{12}

\begin{scope}[scale=0.6]
\foreach \a in {1,...,\n}{
\draw (\a*360 / \n: 2cm) node[point] (p\a) {};
\coordinate (c\a) at (p\a);
}

\begin{scope}[line width = 1pt]
\draw (c1) -- (c11) ;
\draw (c2) -- (c10) ;
\draw (c3) -- (c9) ;
\draw (c4) -- (c8) ;
\draw (c5) -- (c7) ;
\draw (c6) -- (c12) ;
\end{scope}

\begin{scope}[xshift=6cm]
\foreach \a in {1,...,\n}{
\draw (\a*360 / \n: 2cm) node[point] (p\a) {};
\coordinate (c\a) at (p\a);
}

\begin{scope}[line width = 1pt]
\draw (c1) -- (c11) ;
\draw (c2) -- (c10) ;
\draw (c3) -- (c9) ;
\draw (c6) -- (c8) ;
\draw (c5) -- (c7) ;
\draw (c4) -- (c12) ;
\end{scope}

\draw[fill,opacity=0.1,thin] (c1) -- (c2) -- (c3) -- (c4) -- (c9) -- (c10) -- (c11) -- (c12) -- (c1) ;
\draw[fill,opacity=0.1,thin] (c5) -- (c6) -- (c7) -- (c8) -- (c5) ;

\end{scope}

\end{scope}

\node (t) at (1.8,0) {$\leadsto$} ;

\end{tikzpicture}
  \caption{An initial configuration guaranteeing the lower bound of Theorem~\ref{thm:min}, and a possible flip.}
  \label{fig:circle-lower-bound}
\end{figure}

For the lower bound of Theorem~\ref{thm:min}, consider $2n$ points in convex position.
Say, the points are denoted by $p_1$, $p_2$, \dots, $p_{2n}$ in counterclockwise order.
The initial matching links $p_1$ to $p_{n+1}$ and for each $i \in [2,n]$, $p_i$ to $p_{2n+1-i}$ (see Figure~\ref{fig:circle-lower-bound}).

Observe that if the straight-line matching becomes such that there are two disjoint convex sets $C_1$ and $C_2$ whose union contains all the segments of $M$, then the global configuration can be decomposed into two matchings $M_1 \subset C_1$ and $M_2 \subset C_2$. 
Indeed, it is no longer possible that a segment in $C_1$ ever crosses a segment in $C_2$.
And, in particular, $h(M)=h(M_1)+h(M_2)$.
Now, from a configuration of the type of Figure~\ref{fig:circle-lower-bound}, any flip creates two smaller such configurations into two disjoint convex sets.
The process stops when there is a single edge in the configuration. 
Let $H(n)=h(M)$ for a matching $M$ of the type of Figure~\ref{fig:circle-lower-bound} with $n$ segments.
Function $H$ satisfies $H(1)=0$ and $H(n)=H(a)+H(b)$ for any positive integers $a$ and $b$ summing up to $n$.
Therefore, $H(n)=n-1$.   

\begin{figure}[htbp]
  \centering
  \includegraphics{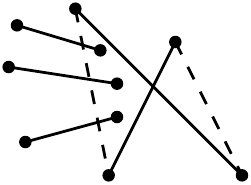}
  \caption{After the depicted flip, the number of crossings goes from $1$ to $3$.}
  \label{fig:MoreCrossings}
\end{figure}
\section{Upper Bounds}

Before we prove the upper bound, observe in Figure~\ref{fig:MoreCrossings} that the number of crossings might increase after a flip.

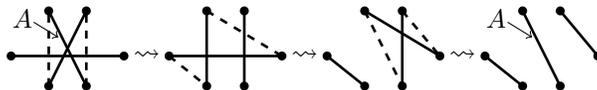
\begin{figure}[htbp]
\centering
\begin{tikzpicture}

\begin{scope}[scale=0.5]

\node[point] (p1) at (0,0.8) {} ;
\node[point] (p2) at (1,0) {} ;
\node[point] (p3) at (1,2) {} ;
\node[point] (p4) at (2,0) {} ;
\node[point] (p5) at (2,2) {} ;
\node[point] (p6) at (3,0.8) {} ;

\foreach \i in {1,...,6}{
\coordinate (c\i) at (p\i) ;
}

\begin{scope}[line width = 1pt]
\draw (c1) -- (c6) ;
\draw (c2) -- (c5) ;
\draw (c3) -- (c4) ;
\draw[dashed] (c2) -- (c3) ;
\draw[dashed] (c4) -- (c5) ;
\end{scope}
\node (t1) at (3.61,0.8) {$\leadsto$} ;

\node (A) at (0.3,1.8) {$A$} ;
\draw[->] (0.6,1.7) -- (1.25,1.3) ;

\begin{scope}[xshift=4.2cm]
\node[point] (p1) at (0,0.8) {} ;
\node[point] (p2) at (1,0) {} ;
\node[point] (p3) at (1,2) {} ;
\node[point] (p4) at (2,0) {} ;
\node[point] (p5) at (2,2) {} ;
\node[point] (p6) at (3,0.8) {} ;

\foreach \i in {1,...,6}{
\coordinate (c\i) at (p\i) ;
}

\begin{scope}[line width = 1pt]
\draw (c1) -- (c6) ;
\draw (c2) -- (c3) ;
\draw (c4) -- (c5) ;
\draw[dashed] (c1) -- (c2) ;
\draw[dashed] (c3) -- (c6) ;
\end{scope}
\node (t2) at (3.61,0.8) {$\leadsto$} ;
\end{scope}

\begin{scope}[xshift=8.4cm]
\node[point] (p1) at (0,0.8) {} ;
\node[point] (p2) at (1,0) {} ;
\node[point] (p3) at (1,2) {} ;
\node[point] (p4) at (2,0) {} ;
\node[point] (p5) at (2,2) {} ;
\node[point] (p6) at (3,0.8) {} ;

\foreach \i in {1,...,6}{
\coordinate (c\i) at (p\i) ;
}

\begin{scope}[line width = 1pt]
\draw (c1) -- (c2) ;
\draw (c3) -- (c6) ;
\draw (c4) -- (c5) ;
\draw[dashed] (c3) -- (c4) ;
\draw[dashed] (c5) -- (c6) ;
\end{scope}
\node (t3) at (3.61,0.8) {$\leadsto$} ;
\end{scope}

\begin{scope}[xshift=12.6cm]
\node[point] (p1) at (0,0.8) {} ;
\node[point] (p2) at (1,0) {} ;
\node[point] (p3) at (1,2) {} ;
\node[point] (p4) at (2,0) {} ;
\node[point] (p5) at (2,2) {} ;
\node[point] (p6) at (3,0.8) {} ;

\foreach \i in {1,...,6}{
\coordinate (c\i) at (p\i) ;
}

\begin{scope}[line width = 1pt]
\draw (c1) -- (c2) ;
\draw (c3) -- (c4) ;
\draw (c5) -- (c6) ;
\end{scope}

\node (A) at (0.3,1.8) {$A$} ;
\draw[->] (0.6,1.7) -- (1.25,1.3) ;
\end{scope}

\end{scope}

\end{tikzpicture}
  \caption{Segment $A$ disappears and reappears.}
  \label{fig:reappearing}
\end{figure}

It is also possible that a segment that has disappeared after a flip reappear after some more flips (see Figure~\ref{fig:reappearing}).
These two observations suggest that there is no straightforward way of getting a good upper bound.

\begin{figure}[b]
  \centering
  a) \includegraphics{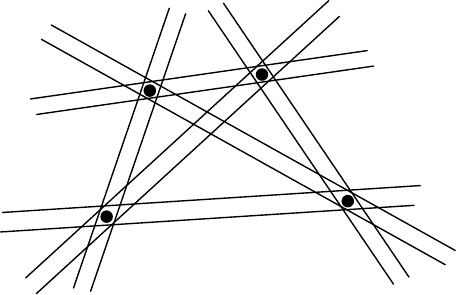} b)  \includegraphics{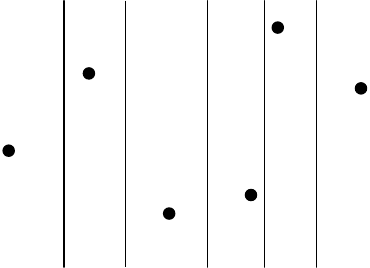}
  \caption{a) Construction of $\mathcal{L}$. b) Construction of $\mathcal{K}$.}
  \label{fig:QuadraticLines}
\end{figure}

For the upper bound of Theorem~\ref{thm:max}, we define a potential function $\Phi_{\mathcal{L}}(M)$ that depends on a well-chosen set of lines $\mathcal{L}$.
We show that $\Phi_{\mathcal{L}}(M) \leq 4 n^3$ and that $\Phi_{\mathcal{L}}$ decreases by at least four after any flip.
The potential function $\Phi_{\mathcal{L}}(M)$ is defined as the number of intersections between a line of $\mathcal{L}$ and a segment of $M$.
We define $\mathcal L$ as follows. Given two points $p,q\in P$ let $\ell$ be the supporting line of $p$ and $q$. We add to $\mathcal L$ the two lines slightly above and below $\ell$ (see Figure~\ref{fig:QuadraticLines}~a)).

\begin{figure}[htbp]
  \centering
  \includegraphics{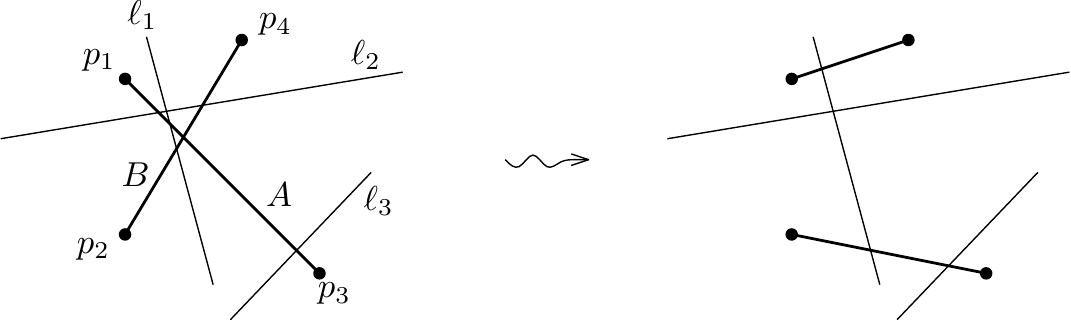}
  \caption{Flipping $A$ and $B$ yields fewer segment-line intersections.}
  \label{fig:LineSegmentCrossings}
\end{figure}

It holds that $|\mathcal{L}|= 2 \binom{2n}{2} \leq 4n^2$. 
As any line and segment can cross at most once it follows $\Phi_{\mathcal{L}}(M) \leq  |\mathcal{L}| \cdot |P| = 4n^3 $.
It remains to show that the number of segment-line intersections decreases by at least four in any flip.
Consider two crossing segments $A$ and $B$ on points $Q = \{ \, p_1,\ldots, p_4 \, \}$ as in Figure~\ref{fig:LineSegmentCrossings}. 
Note that there are only three combinatorial types of lines intersecting the convex hull of $Q$. Either a line separates $p_1$ and $p_2$ from $p_3$ and $p_4$ as $\ell_1$; a line separates $p_1$ and $p_4$ from $p_2$ and $p_3$ as $\ell_2$; or a line separates one point from the other three as $\ell_3$.
For every type of lines the number of intersections does not increase after flipping $A$ and $B$.
It is also easy to see that the number of intersections decreases by two for lines of type $\ell_1$ or $\ell_2$ after flipping $A$ and $B$.
By definition of $\mathcal{L}$, there exists for every crossing of two segments at least two lines of type $\ell_1$ and at least two lines of type $\ell_2$. 
Thus $\Phi_{\mathcal{L}}(M)$ decreases by at least four as claimed.



\begin{figure}[htbp]
  \centering
  \includegraphics{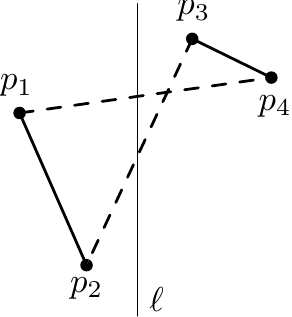}
  \caption{The number of crossings between $\ell$ and the segments of the matchings decreases by $2$. 
}
  \label{fig:MinFlip}
\end{figure}

For the upper bound of Theorem~\ref{thm:min}, we define a different set of lines $\mathcal{K}$ ,which contains one vertical line between any two consecutive points ordered in $x$-direction, see Figure~\ref{fig:QuadraticLines}~b).
It follows $\Phi_{\mathcal{K}}(M) \leq n^2$ since $|\mathcal K|=n-1$. 
We have to show that $\Phi_{\mathcal{K}}$ decreases by at least two after each flip. Let $A$ and $B$ be two crossing segments on the points $p_1,p_2,p_3,p_4$ ordered by $x$-coordinate. Then we replace $A$ and $B$ by $\text{seg}(p_1,p_2)$ and $\text{seg}(p_3,p_4)$, see Figure~\ref{fig:MinFlip}. It is clear that at least one line $\ell$ between $p_2$ and $p_3$ is not crossed after the flip and was crossed twice before the flip.

\paragraph*{Acknowledgments}
The second author thanks the organizers and participants of the Eml\'{e}kt\'{a}bla Workshop in the Summer 2015 for a nice event and interesting discussions respectively. (There are too many to list names and we do not want to forget someone.)
Both authors are supported by the ERC grant PARAMTIGHT: "Parameterized complexity and the search for tight complexity results", no. 280152.

\vspace{1cm}


\end{document}

%% file: commands.tex
\usepackage{comment}
\usepackage{graphicx}

\usepackage[numbers,sort&compress]{natbib}

\usepackage{thmtools}
\usepackage{thm-restate}

\usepackage{amsmath}
\usepackage{amssymb}
\usepackage{xspace}
\usepackage{rotating}
\usepackage[color=green!20]{todonotes}
\newtheorem{theorem}{Theorem}

\usepackage{etoolbox}
\usepackage{aliascnt}

\newcounter{Bew1}
\newcounter{Bew2}
\newcounter{Def1}